# Efficiency of Cathodoluminescence Emission by Nitrogen-Vacancy Color Centers in Nanodiamond

*Huiliang Zhang[1]‡, David R. Glenn[2]‡, Richard Schalek[3,4], Jeff W. Lichtman[3,4], Ronald L. Walsworth[1,2,3]\**

[1]Department of Physics, Harvard University, Cambridge, 02138

[2]Harvard-Smithsonian Center for Astrophysics, Cambridge, MA 02138

[3]Center for Brain Science, Harvard University, Cambridge, 02138

[4]Department of Molecular and Cellular Biology and Center for Brain Science, Harvard University, Cambridge, MA 02138

ABSTRACT: Correlated electron microscopy and cathodoluminescence (CL) imaging using functionalized nanoparticles is a promising nanoscale probe of biological structure and function. Nanodiamonds (NDs) that contain CL-emitting color centers are particularly well suited for such applications. The intensity of CL emission from NDs is determined by a combination of factors, including: particle size; density of color centers; efficiency of energy deposition by electrons passing through the particle; and conversion efficiency from deposited energy to CL emission. We report experiments and numerical simulations that investigate the relative importance of each of these factors in determining CL emission intensity from NDs containing nitrogen-vacancy (NV) color centers. In particular, we find that CL can be detected from NV-doped NDs with dimensions as small as ~ 40 nm, although CL emission decreases significantly for smaller NDs.

The realization of structural imaging with sub-optical resolution in conjunction with nanoscale localization of molecular labels in the same field of view is an important challenge in electron microscopy (EM), with great potential impact in both the life and physical sciences.[1,2] In recent years, various approaches towards this goal have been pursued, including marking distinct epitopes with: (i) different sizes[3] or shapes[4] of nanoparticles (NPs) that provide electron-density contrast; (ii) NPs with distinguishable energy-dispersive X-ray (EDX) contrast[5]; and (iii) NPs with visible wavelength (color)-distinguishable cathodoluminescence (CL) contrast[6]. Among these methods, NP CL is the newest and perhaps the most promising, given its high detection sensitivity[6,7] and the unambiguous spectral separability of such labels.

Cathodoluminescence is the light emitted during relaxation of electronic excitations in a solid, generated by interaction with an electron beam.[2,8,9] CL emission at visible wavelengths (photon energies ~1 eV) has been used extensively as a probe of material properties, such as plasmonics in metallic nanostructures[10,11], polymers[12], and electronic bands of insulators and semiconductors[13]; and in electron microscopy with e-beam energies on the order of a several hundred eV to several hundred keV. The wavelength of CL emission is determined by the electronic energy levels excited [14] and can be controlled to some extent in semiconductors and insulators (such as diamond) by doping or implanting color centers or other defects with well-known emission spectra into the host crystal. It has been shown[6,15] that multi-color nanoscale molecular localization in EM can be achieved by employing several independent types of defect-containing NPs as markers, with each NP type producing spectrally-distinct CL and able to be conjugated to ligands with high affinity for a specific molecular target.

The nitrogen-vacancy (NV) center[16,17] in diamond[18,19] is a particularly important color center for applications in correlative CL and EM imaging, emitting primarily in the red ($\lambda$ ~575 nm – 750 nm) region of the spectrum. The NV center has been widely investigated because of its unique optical and spin properties, and NV centers in nanodiamonds (NDs) have been employed in diverse optical imaging applications, such as: super-resolution imaging by stimulated-emission depletion (STED)[20]; long-term *in vivo* tracking of transplanted stem cells[21]; and local intracellular temperature monitoring[22]. The well-described optical properties of the NV center in NDs make it an ideal system for detailed study of the CL emission in NPs.

For molecular localization applications in EM, markers should be both small, for high spatial resolution and efficient staining in tissue, and bright, for rapid and reliable imaging. In the present work, we investigate the dependence of NV-ND CL emission intensity on a variety of parameters, including e-beam energy and intensity, ND size, and NV number per ND. We present an empirical model describing how these parameters influence CL intensity, and perform a series of correlated scanning electron microscope (SEM), photoluminescence (PL), and atomic force microscope (AFM) measurements to characterize each parameter, with the goal of estimating the minimum size of particles that can be used to generate detectable CL signals. In particular, we study the scaling of CL intensity with ND size, and conclude that surface effects (i.e., either increased rates of carrier recombination at the surface, or quenching of NV fluorescence due to interaction with localized surface states) are not an important factor limiting CL intensity emitted from NDs of size $d > 50$ nm. We observe detectable signals from particles as small as $d \sim 40$ nm, limited primarily by the signal-to-noise ratio (SNR) of our CL imaging setup.

## Results and Discussion

**Parameters Determining CL Intensity:** We first introduce a phenomenological expression for the dependence of emitted CL intensity on electron beam parameters and the size of the probed ND. Electrons impinging on the diamond have an energy-dependent cross section for inelastic interactions with the crystal, resulting in a finite energy deposition efficiency (defined as the ratio of total energy deposited to initial kinetic energy of a single incident e⁻) denoted by $\eta(V,d)$. This quantity depends in general on the accelerating voltage $V$, and the particle size $d$ that determines interaction length of the electron passing through the nanocrystal.

Of the energy inelastically deposited in the crystal by the incident electrons, only a fraction produces electronic excitations in the native NV centers, which then decay radiatively and give rise to detectable CL. This conversion efficiency, here denoted $\xi$, will in general be an average over individual energy conversion efficiencies for each NV center in the particle, which may vary due to details of the local crystal environment such as the density of nearby electron donors (typically substitutional nitrogen defects), surface states, etc. However, the carrier diffusion length[23] in the NDs of interest is comparable to or larger than the particle size (typically 20 - 200 nm in diameter), precluding spatially resolved electron-beam excitation of individual NV centers. We can therefore measure only the integrated CL from the full particle, and must consider the average conversion efficiency $\bar{\xi}$, such that $\xi = N \times \bar{\xi}$, with $N$ the number of NV centers present in the particle. Due to possible saturation effects arising from the finite radiative lifetime of the NV centers, this efficiency can depend on the e-beam current $I$. Furthermore, we conjecture that $\bar{\xi}$ may also depend on the particle size, since smaller particles may have higher rates of carrier recombination at surfaces, resulting in a reduction of the average energy conversion efficiency to CL photons. We parameterize the particle size for this purpose with the linear dimension $d = v_{ND}^{1/3}$, where $v_{ND}$ is the ND volume measured by SEM and AFM. Taking all of these factors into account, we expect that the total observed CL signal intensity $S_{CL}$, integrated over the area (A) and angular acceptance ($\Omega$) of the detection optics, is given by:

$$S_{CL}(V, I, d) = \int s_{CL} dA \, d\Omega \;\propto\; (V \times I) \times \eta(V, d) \times N \times \bar{\xi}(I, d) \tag{1}$$

Here, the integrated CL intensity has units of power, as does the energy flux of the electron beam $(V \times I)$, while the NV number, energy deposition and energy conversion factors are dimensionless. In practice, the electron beam diameter (limited in our experiments to ~4 nm by diffraction in the best case, and more typically to ~20 nm due to defocus and/or sample charging) is much smaller than the NDs, such that the CL image of each particle is distributed over many detection pixels. (Pixel size is determined by electron beam raster parameters, and is chosen to match the e-beam diameter.) In the experiments that follow, we consider only the peak CL intensity, which generally occurs as the beam passes over the center of each ND.

**Characterization of Nanodiamond Shape and Size:** We began by measuring the size distribution and morphology of the NDs under investigation. For the purpose of this study, we looked at He$^+$-irradiated, type Ib NDs, which contain the highest NV concentration of any commercially-available sample currently known to us (see Methods for details). Preliminary analysis by dynamic light scattering (DLS) in aqueous suspension indicated a mean hydrodynamic diameter of 82 nm, with standard deviation 22 nm. We then prepared a sample of NV-containing NDs on a silicon wafer and imaged 20 fields of view in both AFM and SEM (Figure 1a). (Also see Supporting Figure S1 for a complete set of correlated microscopy data.) Measurements were made for a total sample size $n = 257$ particles. Transverse dimensions, $w_x$ and $w_y$ of each particle were obtained from the secondary electron SEM images, while particle heights $h$ were extracted from the AFM measurements. We observed that the NDs tended to be ellipsoidal or plate-like, with the shortest dimension $h$ perpendicular to the substrate surface (Fig. 1b-1c). Unless otherwise indicated, we used the average dimension $d = (w_x \times w_y \times h)^{1/3}$ to parameterize ND size for all further measurements, because (i) this was the most appropriate value to describe achievable CL labeling resolution, and (ii) it was most relevant for estimating the number of NV centers per particle, assuming constant NV density. We note, however, that we repeated most analyses with the alternative parameterization $d_{alt} = h$, since this generally sets the shortest distance between any point on the interior of a given ND and the nearest surface, which may be most relevant for surface-related quenching and recombination effects in small particles. In our measurements and analyses, no significant difference was observed between the two parameterizations. The distribution of $d$ values for our ND sample was peaked at 93 nm, with a standard deviation of 32 nm (Figure 1d).

**Calculated and Measured Energy Deposition Efficiency, $\eta(V, d)$:** We studied the dependence of the deposited energy fraction for incident electrons $\eta$, on the e-beam accelerating voltage $V$, and the size, $d$, of the NDs probed. Because the rate of inelastic energy loss decreases with electron energy ($dE/dz \sim E^{-2/3}$ in the model of Kanaya and Okayama[24]), we expected that highly energetic incident electrons should pass ballistically through the NDs with minimal energy loss, whereas extremely low energy electrons should be completely stopped. Thus, for efficient generation of CL emission in particles of a given size, some intermediate electron energy that gives rise to a stopping range close to the particle size should be optimal. This prediction was supported by preliminary numerical calculations made using the CASINO software package[25,26] (Supporting Figure S2).

To measure $\eta(V,d)$ experimentally, we repeatedly imaged a subset of NDs in CL, while varying the electron accelerating voltage $V$. Assuming the efficiency of conversion from deposited energy to CL intensity $\xi$ remained constant during the measurement for any given particle, the $V$-dependence of the deposition efficiency $\eta$ for that particle could be determined from the observed CL intensity since, from equation (1), $\eta(V,d) \propto S_{CL}/V$.

The measured dependence of energy deposition efficiency on accelerating voltage for several NDs is shown in Figure 2a. All CL images were obtained with an e-beam current of 1 nA, well above the observed saturation current (defined as the current at which observed CL rate reaches half of its large-current asymptotic value) for NDs in the size range of interest (Supporting Figure S3). Because we measured only $S_{CL}$ vs. $V$, while $\xi$ was unknown for any particular particle, the deposition efficiency $\eta$ was determined only to within a constant for each ND. We therefore divided $S_{CL}/V$ by its maximum value $(S_{CL}/V)_{max}$ for each particle to obtain the normalized energy deposition efficiency $\eta(V,d)/\eta_{max}(d)$. As expected, this quantity took on a maximum value at an accelerating voltage on the order of a few keV, which was repeatable for particles of a given size $d$, and increased with increasing $d$. Comparison with CASINO simulations of energy deposition efficiency showed similar behavior as a function of voltage. We note, however, that the CL signal dropped off faster at low $V$ than predicted by the numerical calculations. This discrepancy was likely due to increased surface recombination of carriers generated by low-energy electrons that were stopped immediately upon entering the crystal; thus, at low $V$, the assumption that $\xi$ is independent of $V$ may not be strictly correct. Nevertheless, for ND sizes in the range of 20 nm – 200 nm and $V > 2$ keV, the numerical simulations match the data well enough to justify their use for estimating $\eta(V,d)$ in later experiments. Figure 2b shows the calculated efficiency $\eta(V=5 \text{ keV},d)$ as a function of particle size $d$ and aspect ratio $h/w$, where the model particles were taken to be prolate ellipsoids, $w_x = w_y \equiv w \geq h$. This type of calculation was used in later experiments (with measured ND dimensions as inputs – see section V) to normalize CL measurements with respect to $\eta$, and thereby to determine the conversion efficiency $\xi$.

**Number of NVs per Particle, $N$:** To estimate the number of NV centers present in each ND, we made quantitative PL measurements with a home-built confocal microscope[27]. We first calibrated the system using a separate sample of NDs with low NV density, to determine the expected mean photon count rate for a single NV center. We scanned a circularly-polarized excitation beam at 532 nm with intensity slightly above saturation (~$10^5$ W/cm$^2$) over the sample, split the detected fluorescence into two equal channels, and measured correlations in photon arrival times to isolate NDs containing only a single NV (Figure 3a). Interference filters in the detection path were chosen to admit PL light from NVs in the neutral (NV$^0$) and negative (NV$^-$) charge states, since both produce CL under electron beam excitation[28]. For NDs containing a single emitter, we could distinguish NV charge states by the presence or absence of a bunching shoulder in the g$^{(2)}$ data associated with shelving in the NV$^-$ electronic dark state.[29] However, no correlation was found between the charge state determined by this method and the measured PL count rate, so all NDs containing single emitters were counted together as part of a single distribution (Figure 3b). Over the ~40 NDs probed that contained single NV centers, the mean photon count rate was ~32 kilocounts per second (kcps), and the standard deviation was 9 kcps. While broad, this distribution still allowed precise estimation of NV number for NDs with many NVs. (For example, in the limit of uncorrelated single-NV count rate distributions, a total count rate of 320 kcps could be attributed to the presence of 9-11 NVs with 95% probability.)

Finally, we collected correlated PL images of the set of NDs characterized in section II above, using the same optical excitation intensity and detection filters as for the single-NV PL calibration. This allowed us to estimate the NV number in each particle, and thereby to determine the NV density for each particle (Figure 3c). The observed PL rates indicate a constant average NV density of about $2 \times 10^{-4}$ /nm$^3$ for NDs larger than $d \sim 100$ nm, but significantly lower for smaller NDs. The observed dependence of NV density on $d$ is not consistent with an excluded volume model, in which the effective dimensions of NDs are reduced to account for surface quenching or overestimation of $w_x$ and $w_y$ due to SE image defocus (Supporting Figure S4). We therefore conclude that NV formation was suppressed during sample preparation for small NDs, likely because of some combination of reduced efficiency of vacancy generation by energetic He$^+$ ions and rapid migration of vacancies to nearby surfaces during annealing. For the 40 keV He$^+$ implantation protocol used to generate vacancies in our numerical simulation using the SRIM software package[30] predicts that the most efficient depth for vacancy generation in bulk diamond is ~150 nm, larger than the size of most of the NDs.

**Measured CL Intensity Distribution and Average Energy Conversion Efficiency, $\bar{\xi}(I, d)$:** We obtained CL images of the same NDs, co-localized with the AFM and PL fields of view described above, to study the efficiency of conversion from deposited e-beam energy to detectable CL. The CL data were collected with an e-beam current $I = 1$ nA, at constant accelerating voltage $V = 5$kV. Analysis of the CL images reveals that, for NDs with nonzero PL signals, the fraction that produce observable CL decreased with particle size for $d < 100$ nm (Figure 4a, inset). The smallest NDs detectable by CL in our system were in the range of 30-40 nm, although only about 10% of the particles in this size range could be distinguished from the noise floor at the 5σ level.

To inform future applications in correlative CL imaging, we wished to determine whether this limited detection rate for small NDs known to contain one or more NV emitters was due to (i) the low CL brightness per NV making it difficult to distinguish from background the signal from NDs containing only a few NVs, or (ii) some additional size-dependent effect such as CL quenching or surface carrier recombination in small NDs suppressing CL emission per NV below the level observed in larger NDs. Therefore, by direct comparison with PL images, we measured the CL signal intensity as a function of the number of NV emitters in each ND (Figure 4a). We found that the dependence was linear down to the noise floor of the CL images, which became dominant at a CL intensity corresponding to ~20 NV centers. CL emission by the Si wafer set the background level, approximately an order of magnitude larger than the detector dark count rate. Because tissue embedded in resin for EM imaging (and likely most other biological or material samples of interest) produce CL emission rates comparable to or greater than the Si wafer[1], our results suggest that a relatively high density of NV centers is required for ND-based correlative imaging with sub-50 nm spatial resolution under the present imaging conditions.

Finally, to investigate the average energy conversion efficiency $\bar{\xi}$ and test for intrinsic ND size-dependence beyond the number of NV centers per particle, we calculated the energy deposition efficiency $\eta$. Particle dimensions from the AFM images were used as inputs to CASINO to numerically calculate $\eta$ for each ND. Dividing $S_{CL}$ for each particle by $N$ and the calculated $\eta$ yielded a "corrected" CL signal, proportional to $\bar{\xi}$, which we then plotted as a function of particle size (Figure 4b). We included only NDs with CL brightness distinguishable from background for this analysis. For this subpopulation ($n = 195$) of NDs, we found no evidence of decreasing average energy conversion efficiency $\bar{\xi}$ at small $d$. Thus, CL quenching and surface carrier recombination are not likely to be limiting factors for correlative CL imaging using currently available NDs.

Contrary to expectation, however, we observed a rapid decrease in energy conversion efficiency $\bar{\xi}$ (going approximately as $\sim d^3$) for NDs larger than $\sim 80$ nm. We attributed the decrease in $\bar{\xi}$ for the largest NDs to short carrier diffusion lengths, known to be associated with high N impurity concentration[23], which effectively limited the instantaneous volume over which NV centers could be excited by the SEM beam at any given scan position. To test this hypothesis, we made EPR measurements (Supporting Figure S5) on macroscopic ensembles of the NDs used in this work. These measurements indicated a N concentration on the order of 200 ppm, which should limit the carrier diffusion length to much less than $L_D < 500$ nm[23].

**Size Dependence of Nanodiamond CL Spectra:** For applications in multi-color correlative microscopy, spectral distinguishability is another important factor in the selection of suitable nanoparticle labels. We therefore collected CL spectra for NV-containing NDs of varying sizes (Figure 5a). We observed that for decreasing ND size (parameterized in these measurements by the largest transverse dimension, $w$, since the TEM grid substrates used for these measurements prevented acquisition of correlated AFM data), the decrease in red CL emission due to reduced NV number was accompanied by a higher fraction of CL emitted in the blue, associated with so-called A-band defects[31]. Interestingly, the absolute intensity of blue CL emission also increased in the smaller particles (Figure 5b), indicating a much higher density of physical defects such as dislocations and twinning in the smaller particles. This does not pose a fundamental problem for multi-color correlative imaging, since blue CL-emitting nanoparticle labels have been demonstrated with nearly zero emission in the red, thus allowing a very low threshold in the red channel to be used to distinguish between the two species[6]. Nevertheless, it may be important to consider the spectral purity of CL emission by NV-containing NDs when attempting to construct new sets of orthogonal labels.

**Conclusion:** We conducted a thorough study of the cathodoluminescence (CL) emission properties of nanodiamonds (NDs) containing a high concentration of nitrogen-vacancy (NV) color centers for use in correlative microscopy. Because minimizing the size of the ND labels is crucial for practical applications, we attempted to isolate the mechanisms by which particle size affects CL brightness. We first carried out correlated AFM and CL imaging on a small sample of NDs, varying the electron acceleration voltage and particle size, and compared these measurements with numerical calculations made using the CASINO software to calibrate the fraction of energy deposited by incident electrons. We then performed correlated AFM, PL and CL imaging on a larger sample which, in conjunction with the calibrated CASINO calculations, allowed us to extract the number of NV centers per ND and the efficiency of conversion from deposited $e^-$ energy to CL emission for each ND. We found that this conversion efficiency increases rapidly with decreasing particle size down to approximately 80 nm, likely because the short carrier diffusion length in type Ib diamond prevents excitation of all NVs in the larger NDs. No decrease in conversion efficiency was observed for very small particles. We detected CL from NDs as small as ~40 nm, although only the brightest particles in this size range could be seen reliably above the CL background of our substrate. Finally, we found that the ratio of A-band (blue emission) to NV (red emission) CL increases dramatically for particles smaller than 100 nm, although this is not an important practical limitation for correlative microscopy with appropriate detection filters and image thresholding. Given the average NV density observed in this sample, we expect that increased averaging time will allow reliable detection of a high fraction of particles in the ~30-40 nm range. Operation at lower accelerating voltages *V* could also be a viable strategy to improve the relative energy deposition efficiency η for small particles while reducing background rates, although care must be taken to avoid resolution degradation due to sample charging in this case. With improvements in ND fabrication, CL imaging of even smaller labels will likely become possible. Furthermore, while the present work has focused on NV centers as a model system with well-quantified bulk PL and CL properties, we have previously identified brighter nanoparticle CL emitters (*e.g.,* NDs optimized for A-band emission or Ce:LuAG nanophosphors)[6], which provide a promising alternative route to very high resolution CL labeling. Taken together, the results described here demonstrate that NDs containing NVs and other color centers are a promising marker for use in correlative microscopy.

## Methods

**Nanodiamond sample preparation:** NDs were obtained from Academia Sinica, Taiwan[32]. The sample consisted of type 1b NDs with mean diameter ~100 nm, and nitrogen concentration >100 ppm. The suppliers implanted these NDs with 40-keV He$^+$ ions at a dose of ~ $1\times10^{13}$ ions/cm$^2$, then annealed at 450 C for 4 hours to promote vacancy formation[32]. We suspended the NDs in DI water, diluted to 0.01 - 0.001 mg/ml, ultrasonicated, and then drop-cast onto silicon wafer substrates. The substrates were first prepared by inscribing a series of 10 µm grids for coarse image registration, using photo-lithography (Karl Suss MJB4 Mask Aligner) and reactive ion etching (South Bay RIE-2000).

**Optical imaging:** A home-built confocal microscope was used for PL characterization. ND fluorescence was excited using a Nd:YAG laser at 532 nm (intensity ~$10^5$ W/cm$^2$) and detected with an avalanche photodiode (APD, Perkin Elmer SPCM-AQRH-14). Two cascaded 532 nm long pass filters (Semrock) before the APD blocked excitation light while admitting NV emission (for both neutral and negative charge state). For measurements of photon statistics to identify single NV emitters, a 50-50 beamsplitter and a second APD were added to the detection path, and relative photon arrival times between the two channels were recorded using a time correlated single photon counting module (Picoquant Picoharp 300). Autocorrelation time traces were fit according to the function $g^{(2)}(t) = 1 - (1+a)e^{-|t|/\tau_1} + a\, e^{-|t|/\tau_2}$.

**Secondary electron imaging and CL imaging:** CL properties were investigated using a field emission SEM (JEOL JSM-7001F) outfitted with a spectrally-selective, PMT-based CL detection system[6]. Secondary electron images were used to determine particle lateral dimensions. An accelerating voltage of 5 keV was used in most experiments, unless otherwise specified.

**Atomic force microscopy:** An Asylum MFP-3D was used to measure the height of each ND in the direction normal to the substrate surface. An Olympus AC240TS AFM probe provided relatively high hardness and low lever spring constant, such that wear could be minimized during measurements.

**Image co-localization:** By using the fabricated silicon wafers with series of 10 μm grids, we are able to locate the same position for PL, CL, SEI, and AFM imaging modalities (as shown in Figure S2). Data were collected from correlated images of a total of 20 independent 10 μm × 10 μm fields of view. Custom Matlab software was used to carry out image co-localization, and 2D Gaussian fitting was performed to extract particle dimensions, as well as peak PL and CL signal intensities.

**CASINO simulation:** The CASINO (monte CArlo SImulation of electroN trajectory in sOlids) software package[25,26] was used for modeling electron–sample interactions in the scanning electron microscope. Complete electron trajectories were simulated in the sample, allowing modeling secondary electrons, backscattered electrons, absorbed energy, CL emission, X-ray emissions, and other relevant phenomena. In our work, absorbed/deposited energy information was extracted from the simulation. Each calculation was carried out with $5 \times 10^4$ simulated incident electron events, using a 5 nm diameter electron beam centered on an ellipsoidal model ND. The simulated accelerating voltage was 5 kV, unless otherwise specified.

**Nitrogen concentration measurements:** To determine ND nitrogen impurity concentrations, we performed electron paramagnetic resonance (EPR) measurements (Bruker ElexSys E500 EPR) on a 50 μL sample of NDs, with concentration 1 mg/mL. NDs were dispersed in water and then transferred into a quartz EPR sample tube (2 mm thin wall, 100 mm long, Wilmad Glass Co INC, 704-PQ-100M). (2,2,6,6-Tetramethylpiperidin-1-yl)oxyl (TEMPO, 20 μM) was used as calibration standard to estimate nitrogen concentration. This measurement indicated an average N density of ~230 ppm, for our ND sample.

FIGURES

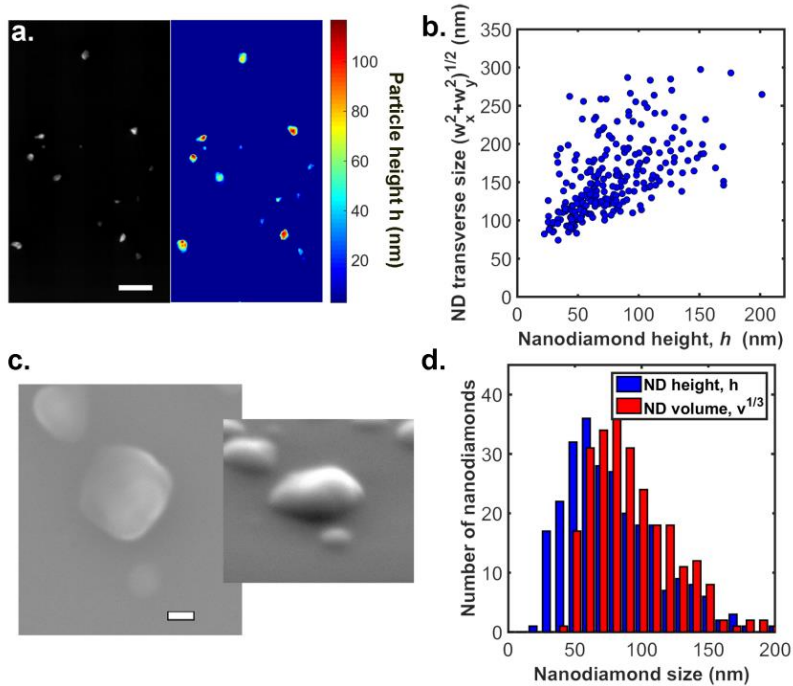

**Figure 1**. Characterization of nanodiamond (ND) size and morphology. **a.** NDs, dispersed on Si wafers, were imaged using correlated secondary electron imaging (SE, left panel) and atomic force microscopy (AFM, right panel) to determine their transverse ($w_x$, $w_y$) and longitudinal ($h$) dimensions, respectively. SE provides a better estimate of transverse dimensions than AFM due to blurring associated with finite tip radius. SE scale bar is 1 μm. **b.** ND aspect ratios for all particles (n=257) in our sample. Most particles were found to be plate-like, with aspect ratios $w/h$ between 1.5 and 3. **c.** Zoomed-in SE images of several NDs from our sample on a tilted stage, illustrating the plate-like morphology of the particles. Left panel shows dimensions of NDs in the plane parallel to the silicon wafer (i.e., $w_x$ and $w_y$); right panel shows the same field of view tilted 60° toward the normal. Scale bar is 100 nm. **d.** Histogram of particle sizes, $d$, calculated using both the shortest dimension $h$ and the geometric average $v^{1/3} = (w_x \cdot w_y \cdot h)^{1/3}$. We use the latter definition in all subsequent measurements.

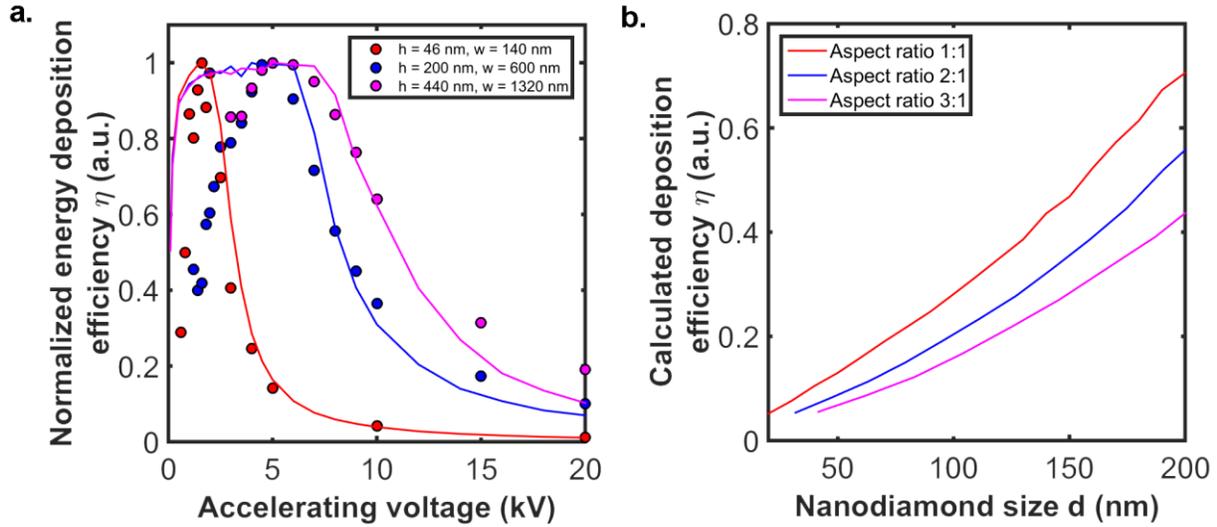

**Figure 2:** Calculation of energy deposition efficiency, $\eta$, and comparison with measurements. **a.** Filled Dots: Normalized energy deposition efficiency $\eta/\eta_{max}$, obtained by recording cathodoluminescence (CL) emission intensity while varying electron beam accelerating voltage $V$. Data shown are for three particles with widely varying size parameters. Solid lines: CASINO calculations of normalized energy deposition efficiency, for model particles of the same dimensions. The calculated values agree well with observation for the parameter range of interest ($V = 5$kV, $d = 20 – 200$ nm), but deviate at low $V$, likely due to surface carrier recombination. **b.** CASINO calculations of absolute deposition efficiency $\eta$ for a range of particle sizes $d$ and aspect ratios $h / w$, for $w = w_x = w_y$, the ND transverse dimensions, and $h$, the ND height along the direction of e-beam propagation. Calculations of this type particles were later used to normalize CL measurements to obtain CL emission efficiency per eV of deposited electron energy.

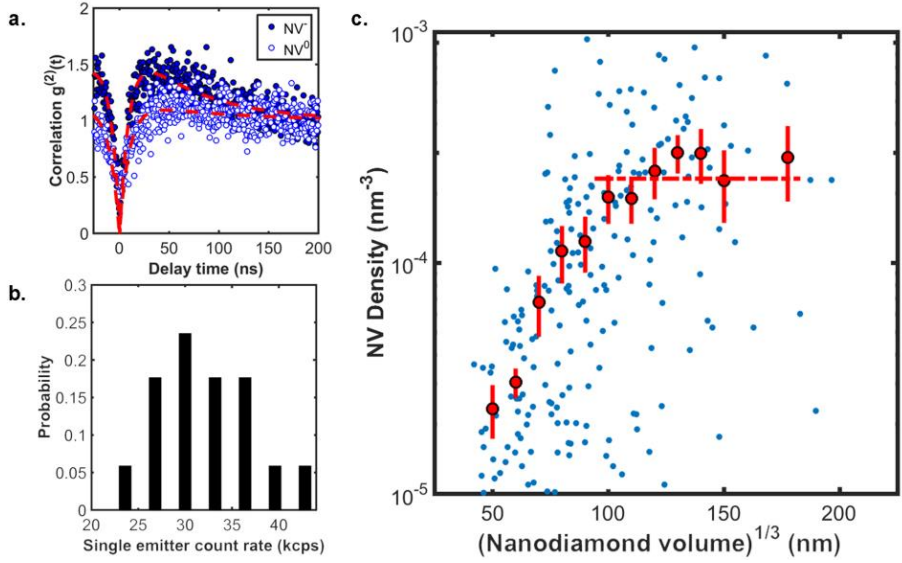

**Figure 3:** Photoluminescence (PL) counting of NV centers. **a.** Characteristic second-order autocorrelation time trace $g^{(2)}(t)$ used to identify NDs containing a single photon emitter (solid blue dots NV$^-$, open blue dots NV$^0$, and red dashed lines fits to experimental data)[22]. NDs containing >1 emitter would show nonzero fluorescence at zero delay. **b.** PL intensity calibration of our confocal microscope, carried out using only NDs containing single NV centers. The distribution of observed count rates per NV was assembled from $n$=23 observations of single-emitter NDs. The number of NV centers in a brighter ND can by estimated by dividing its PL count rate by the mean of this distribution, $\langle S_{PL} \rangle$= 32 kcps. **c.** Density of NV centers for each ND in our main sample ($n$=257), obtained by dividing measured NV number (using calibrated PL) by ND volume (from AFM and SEI images above). Blue points show raw data for each ND, red circles are binned averages. Error bars are the standard error (S.D.) in the mean of each bin, $\sigma/N^{1/2}$. Data are consistent with a constant average NV density of $(2.3 \pm 0.2$ S.D.$) \times 10^{-4}$ nm$^{-3}$ for NDs in the size range $d \geq 100$ nm, indicated by the dashed red line. This density falls off rapidly for smaller NDs, possibly due to decreased efficiency of NV formation.

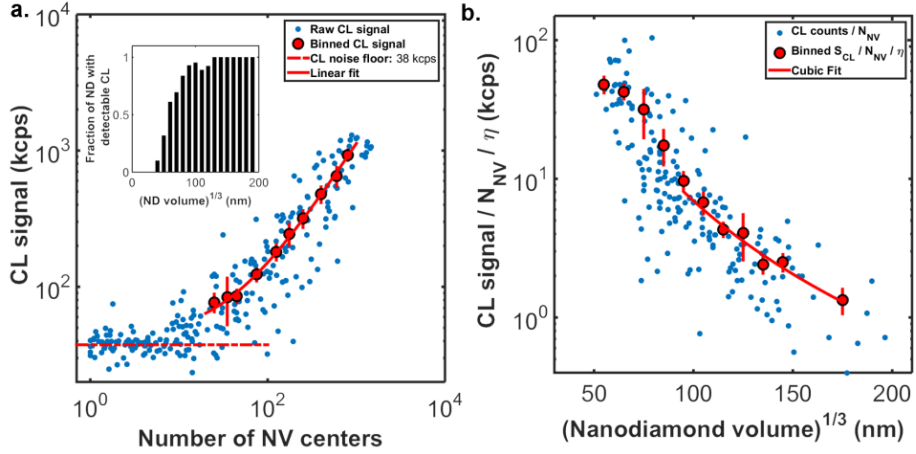

**Figure 4:** Effect of ND size and NV number on CL intensity. **a.** Measured CL signal as a function of number of NV centers (determined by PL intensity) for each ND sampled. Blue points show CL signal for all NDs; red circles are averages binned by NV number. The observed dependence is linear. [Solid red line shows a fit of the form $S_{CL} = c_1 N_{NV} + c_2$, with slope $c_1 = (1.09 \pm 0.03$ S.D.$)$ kcps / NV center]. The smallest number of NV centers detectable is on the order of 10-20, before background noise (red dotted line) becomes dominant. This suggests that NV density will also determine the size limit for CL detection. Inset shows fraction of NV-containing NDs (i.e., with detectable PL signal) that also had detectable CL (at least $5\sigma$ above background noise) as a function of ND size. **b.** ND size-dependence of CL signal per NV center, normalized by calculated $\eta$ to account for size-dependence of energy deposited by the electron beam. The data show no evidence of decreasing CL emission per NV due to surface effects at the smallest measured ND sizes, $d \leq 80$ nm. For large NDs in the size range $d \geq 100$ nm, the CL rate per NV decreases as $\sim d^3$. [Solid red line shows a fit of the form $S_{CL}/N_{NV}/\eta = c_1 (V_{ND}^{1/3})^{-3}$, with $c_1 = (6.9 \pm 0.4$ S.D.$) \times 10^6$ kcps nm$^3$.] We attribute this to finite carrier diffusion lengths due to high N impurity concentrations, which prevent electron hole pairs from recombining at NV centers distant from the electron beam.

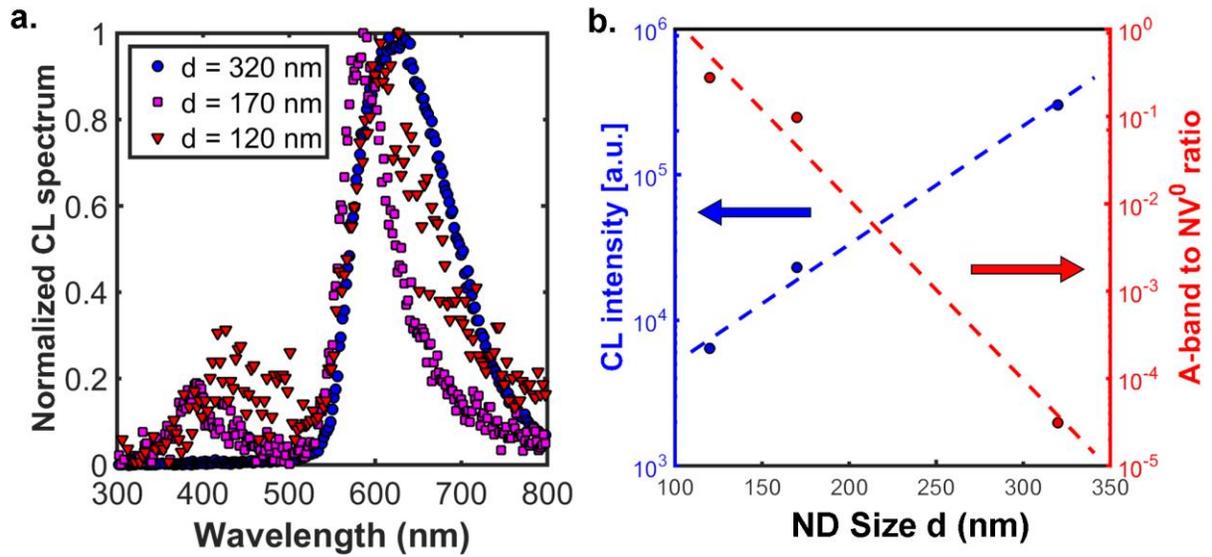

**Figure 5.** Effect of ND size on CL emission spectrum. **a.** CL spectrum for 3 different sizes of ND. Data are normalized to the peak value in each spectrum. Smaller NDs exhibit increased relative emission rates around λ = 400 nm. **b.** Effect of ND size on spectral distinguishability of NDs. Blue points represent total CL signal integrated over the full visible wavelength spectrum. Red points are the ratio of CL signal from A-band (integrated from 300 nm to 470 nm) and from NV (integrated from 550 nm to 750 nm).

## ASSOCIATED CONTENT

**Supporting Information Available:** (i) Typical co-registered data set including SEI, AFM, PL and CL images of NDs; (ii) sample CASINO calculation showing scattering trajectories of electrons by a ND; (iii) characterization of CL intensity saturation with e-beam current; (iv) comparison of models for NV defect number vs. ND size; (v) ND EPR spectra for estimation of paramagnetic N concentration. This material is available free of charge via the Internet at http://pubs.acs.org.

## AUTHOR INFORMATION

**Corresponding Author**

*Corresponding author: Ronald L. Walsworth, rwalsworth@cfa.harvard.edu.

**Author Contributions**

The manuscript was written through contributions of all authors. All authors have given approval to the final version of the manuscript. ‡These authors contributed equally.

**Funding Sources**

This research was funded by the NSF and DARPA.

ACKNOWLEDGMENT: We thank X. Zhou, S. Gradecak, and N. Kasthuri for fruitful discussion and helpful technical assistance.

**Supporting Information**

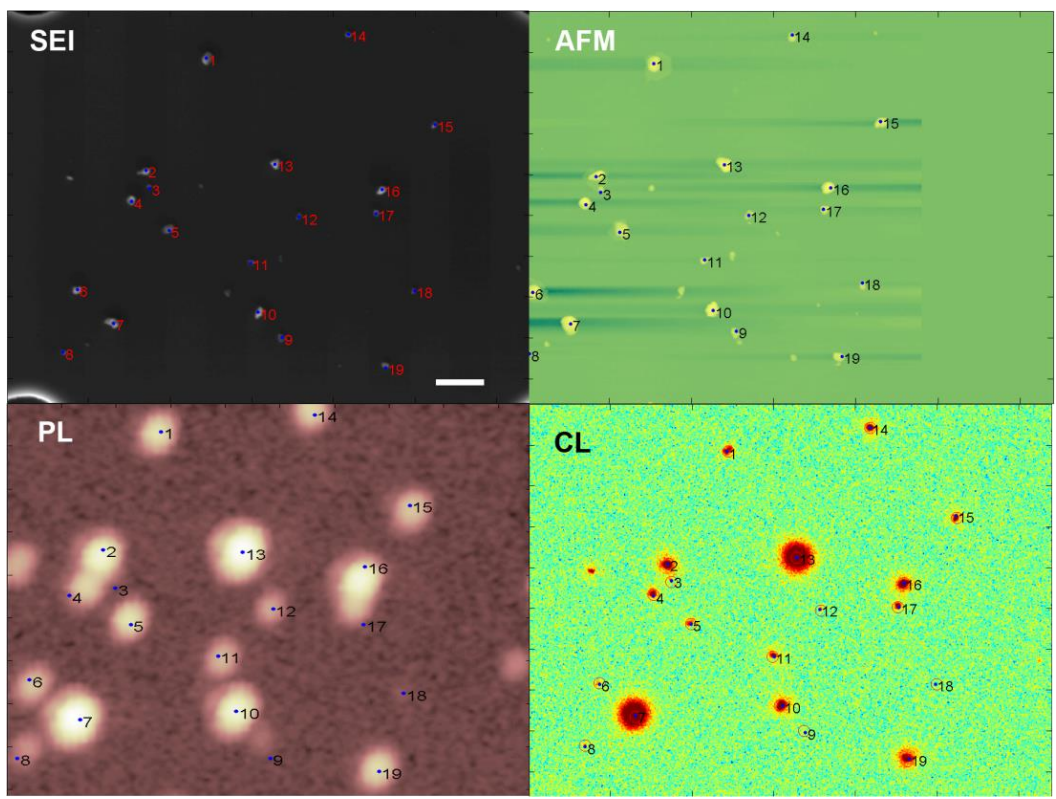

**Figure S1**. Typical co-registered nanodiamond (ND) images. All NDs in our data set were successively imaged using (i) electron microscopy, including simultaneously-detected secondary electron images (SEI) and cathodoluminesce (CL); (ii) atomic force microscopy (AFM); and (iii) photo-luminescence (PL) microscopy. The NDs were deposited onto silicon wafers etched with a numbered grid to facilitate registration of fields of view obtained in different instruments. (The edges of features demarcating one grid location are visible in the corners of the SEI image here.) Image registration was carried out using custom *Matlab* software, and only NDs with visible and non-saturated images in all four instruments were included in the full data set. NDs that were imaged in SEI, AFM, and PL, but had no detectable CL signal, were also included in the data used to generate Figures 1 and 3 of the main text, but were excluded from Figure 4. In the images above, 17 of the 19 numbered NDs were visible in all four images, while two (labeled 9 and 18) were not detectable in CL. The total data set consisted of $n = 257$ co-registered NDs, obtained from 20 sets of images. Of these, $n_{CL>0} = 195$ had CL signals distinguishable from background. Scale bar 1 µm.

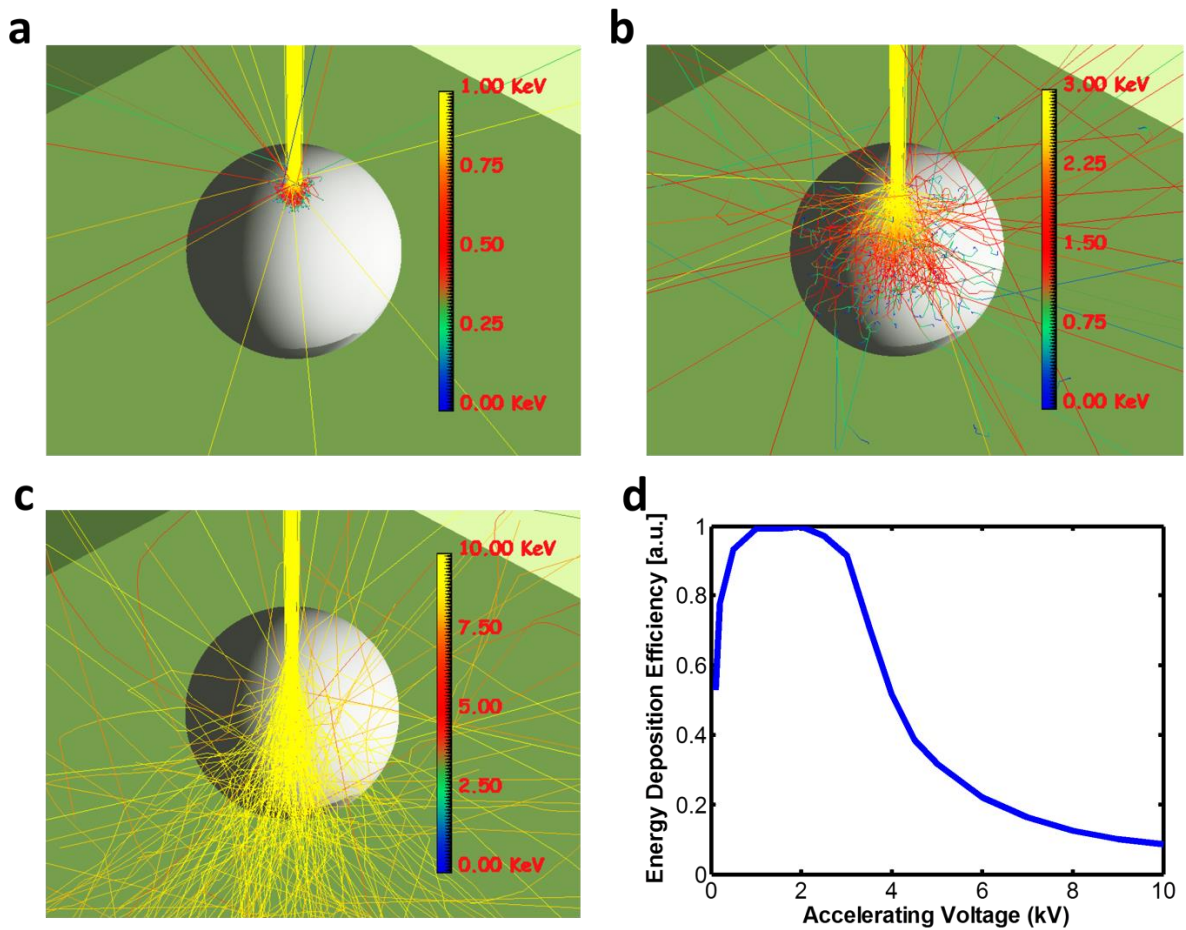

**Figure S2:** Simulation of injected e-beam scattering trajectories. To estimate the efficiency of energy deposition by electrons impinging on nanodiamonds (NDs) of various sizes, we carried out a series of numerical simulations using the CASINO software package. (CASINO = monte CArlo SImulation of electroN trajectory in sOlids). Initial calculations were carried out spherical diamond nanoparticles of diameter $d = 100$ nm, at the following energies: **a**, 1 keV **b**, 3 keV **c**, 10 keV. The color of the electron trajectories in each figure indicates the energy of a particular electron in the calculated ensemble. Electrons undergo continuous energy loss by inelastic scattering, punctuated by random elastic scattering events which result in sudden changes in energy and momentum. In these simulations, slow (1 keV) electrons are completely stopped within the first ~10 nm of the particle, whereas fast (10 keV) electrons travel ballistically through the whole ND while depositing only a small fraction of their energy. **d**, Energy deposition efficiency (ratio of deposited energy to total incident electron energy) as a function of accelerating voltage. For the 100 nm diameter particles, optimized energy deposition efficiency can be achieved for accelerating voltages of about 1 - 3 kV.

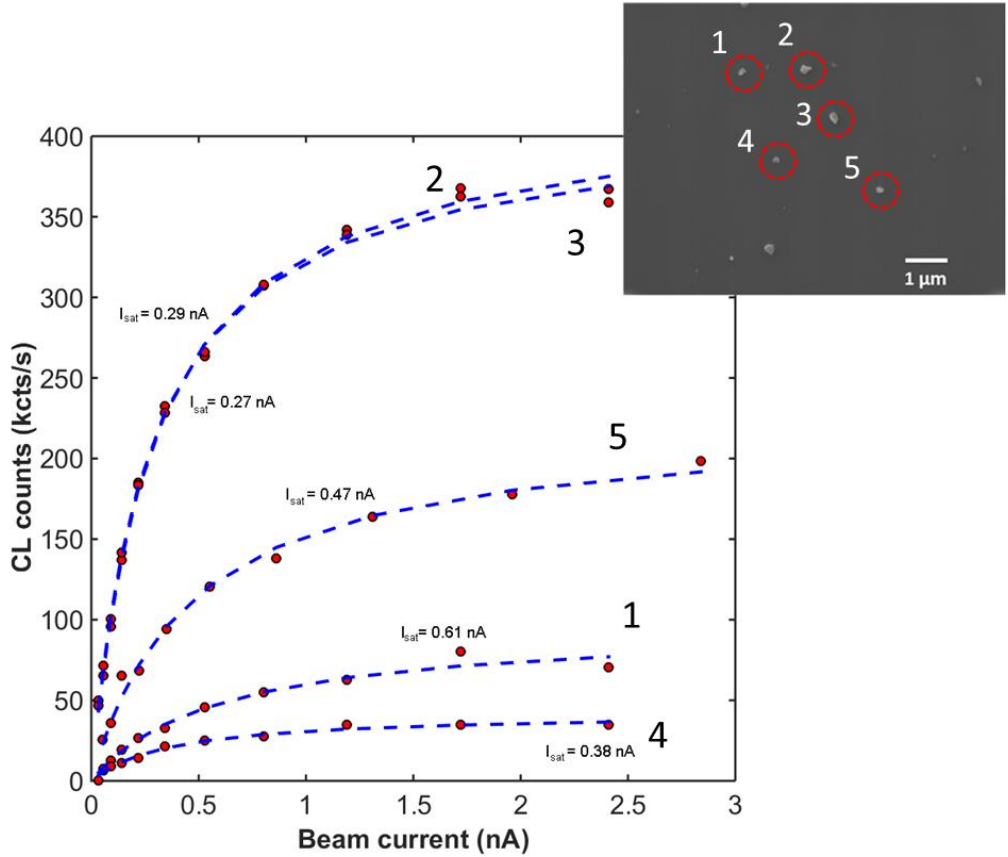

**Figure S3.** Characterization of CL intensity saturation with e-beam current. Inset SEM image shows five typical nanodiamonds (NDs) from our sample. For each of these NDs, we measured the dependence of CL intensity on e-beam current at fixed accelerating voltage V=5 keV. The detected CL intensity saturates at large currents, as the rate of electron-hole-pair (EHP) generation and conversion to nitrogen-vacancy (NV) electronic excitations approaches the NV excited-state lifetime. The observed saturation behavior may also be due in part to charging effects at the sample surface. Typical saturation currents for NDs in our sample were on the order of $I_{sat} \approx 0.3 - 0.5$ nA, obtained by fitting CL data to the following form:

$$S_{CL} = \frac{S_{max}\left(\frac{I}{I_{sat}}\right)}{1+\left(\frac{I}{I_{sat}}\right)}.$$

We performed all CL measurements described in the main text at a current $I = 1$ nA, well above $I_{sat}$, to ensure that variations in the detected CL rates were primarily due to NV density and/or surface quenching-related effects, rather than e-beam-limited EHP generation rates. Higher operating currents could not easily be used due to sample charging issues. Particle sizes, $d = (w_x w_y)^{1/2}$, are 135 nm, 201 nm, 209 nm, 121.4 nm, 135.1 nm for particles 1~5, respectively.

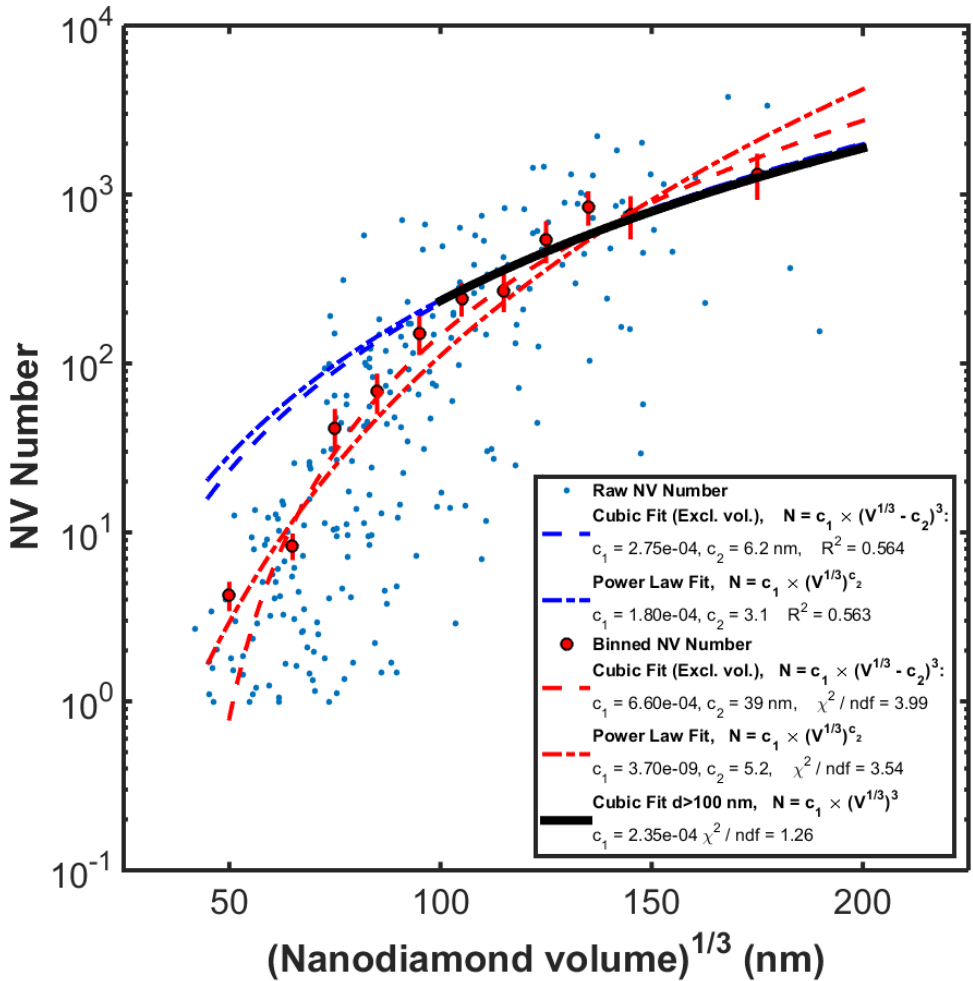

**Figure S4:** Comparison of models for nitrogen-vacancy (NV) defect number vs. nanodiamond (ND) size. Blue dots show estimated number of NV centers per ND (obtained from photoluminescence measurements) as a function of ND size (obtained from secondary electron imaging and atomic force microscopy). ND size is parameterized as the cube-root of the measured ND volume, $d = V^{1/3} = (w_x \cdot w_y \cdot h)^{1/3}$. Red circles show the same data, with NVs binned by size in 10 nm intervals. Error bars are the standard error in the mean of each bin, $\sigma/n^{1/2}$. Fits were made to two different models to investigate dependence of NV number on ND size, including (i) a simple power law $N = (c_1 \cdot V^{1/3})^{c_2}$, and (ii) an excluded volume model, $N = c_1 \cdot (V^{1/3} - c_2)^3$. The power law should yield a $d^3$ dependence in the simplest case of constant NV density. The excluded volume model *assumes* this cubic dependence, but allows for a depleted surface layer of variable thickness, in which NVs are unable to form or have suppressed photoluminescence due to surface quenching. Dashed and dotted blue lines show equal-weighted least squares fits of the un-binned data to each model. The fits are dominated by population of NDs with large $d$, demonstrating the need for binning and appropriate weighting for correct analysis of the data. Dashed and dotted red lines show fits of the binned data to the cubic excluded volume model and the power law model, respectively. Fits on the full data set are consistent with neither model; in both cases, $\chi^2$ / n.d.f.

(number of degrees of freedom) > 3. However, if only the large-volume NDs are included, $d > 100$ nm, then the binned data are consistent with a fit to a simple cubic, $N = c_1 \cdot d^3$ (solid black line). Because the anomalously low NV numbers for small NDs are not consistent with the excluded volume model, we attribute the deviation from constant density to a size-dependent decrease in NV creation efficiency during sample preparation.

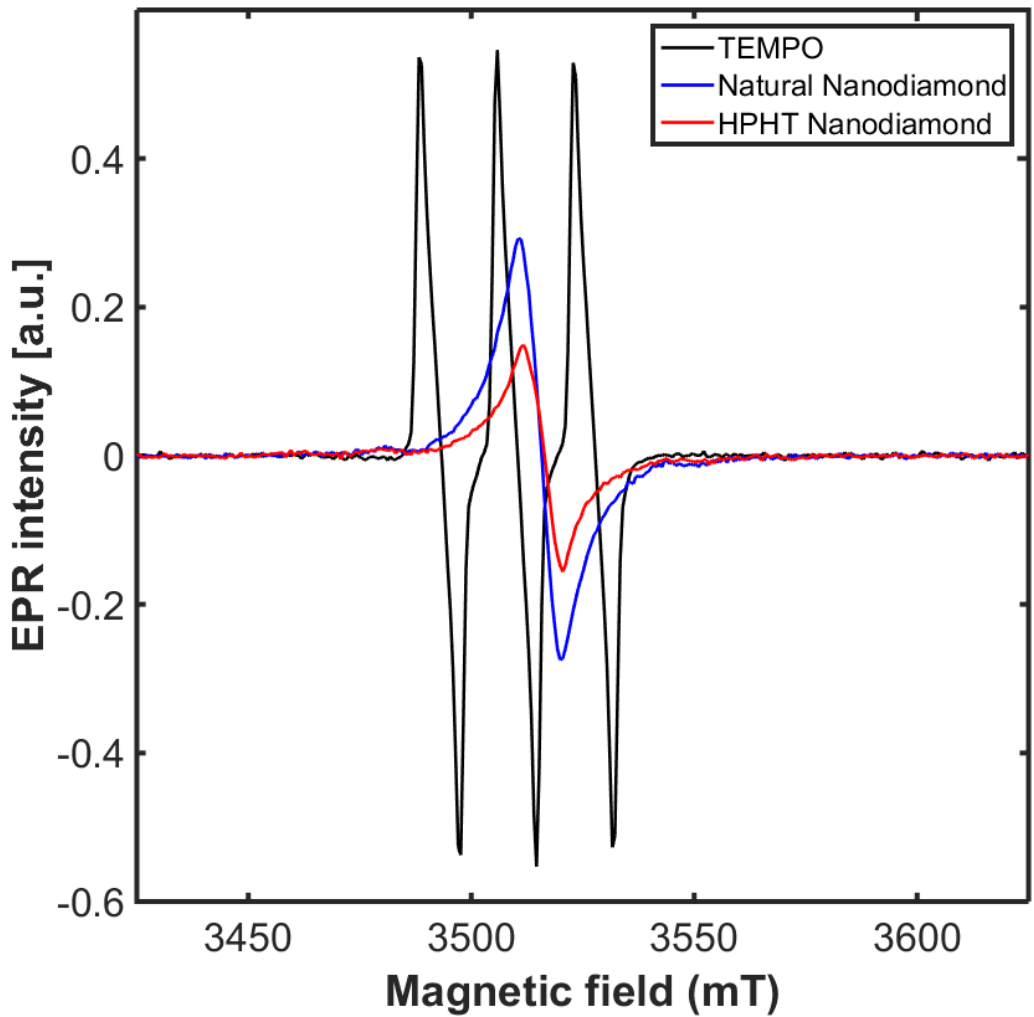

**Figure S6**. Background-subtracted EPR spectra: TEMPO calibration standard (black trace); an auxiliary sample of blue-CL emitting natural nanodiamonds (blue trace); and our primary, red-CL emitting, $He^+$-irradiated HPHT nanodiamond sample (red trace). The irradiated HPHT diamonds were used in all other measurements in this work. To estimate the electronic spin density in each sample, we integrated each spectrum twice (since EPR spectra as shown are the derivative of microwave absorption spectra). The relative doubly-integrated signals were approximately 103 (TEMPO) : 127 (natural ND) : 69 (HPHT ND). Both diamond solutions were 35 μL in volume, prepared at a concentration of 1 mg / mL, while the TEMPO solution was 50 μL in volume and had a concentration of 20 μM. We therefore estimate the paramagnetic (nitrogen) spin concentration in the diamond samples to be ~430 ppm for the natural NDs, and ~230 ppm for the HPHT.